\documentclass[fleqn,usenatbib]{mnras}

\usepackage{amsfonts,amsmath,amssymb,bm,booktabs,cleveref,dcolumn,enumitem,graphicx,multirow,nomencl,physics,xcolor,arydshln,float,balance,orcidlink}

\usepackage[version=4]{mhchem} 

\usepackage[skip=0pt]{caption}

\graphicspath{{Figures/}}

\def\a0{{$a_{\rm 0}$}}

\makenomenclature
\newcolumntype{d}{D{.}{.}{-1}}
\newcolumntype{H}{>{\setbox0=\hbox\bgroup}c<{\egroup}@{}}

\newcommand{\Abinitio}{\emph{Ab initio}}
\newcommand{\AbInitio}{\emph{Ab Initio}}
\newcommand{\abinitio}{\emph{ab initio}}

\newcommand{\etal}{\emph{et al.}}
\newcommand{\cm}{cm$^{-1}$}

\newcommand{\Duo}{Duo}
\newcommand{\DUO}{Duo}
\newcommand{\STATES}{.states} %\texttt
\newcommand{\TRANS}{.trans}
\newcommand{\MODEL}{.model}
\newcommand{\EXOCROSS}{ExoCross}
\newcommand{\EXOMOL}{ExoMol}

\newcommand{\Marvel}{MARVEL}
\newcommand{\MARVEL}{MARVEL}
\newcommand{\Mollist}{MoLLIST}
\newcommand{\MOLLIST}{MoLLIST}
\newcommand{\Molpro}{Molpro}
\newcommand{\MOLPRO}{Molpro}

\newcommand{\LLname}{{\sc ZorrO}}

\newcommand{\ZrObold}{$^\mathbf{90}$Zr$^\mathbf{16}$O}
\newcommand{\ZrOl}{ZrO}

\newcommand{\ZrO}{$^{90}$Zr$^{16}$O}
\newcommand{\isoa}{$^{91}$Zr$^{16}$O}
\newcommand{\isob}{$^{92}$Zr$^{16}$O}
\newcommand{\isoc}{$^{93}$Zr$^{16}$O}
\newcommand{\isod}{$^{94}$Zr$^{16}$O}
\newcommand{\isoe}{$^{96}$Zr$^{16}$O}

\newcommand{\ZrOA}{A~$^1\Delta$}
\newcommand{\ZrOaa}{a~$^3\Delta$}
\newcommand{\ZrOB}{B~$^1\Pi$}
\newcommand{\ZrObb}{b~$^3\Pi$}
\newcommand{\ZrOC}{C~$^1\Sigma^+$}
\newcommand{\ZrOcc}{c~$^3\Sigma^-$}
\newcommand{\ZrOD}{D~$^1\Gamma$}
\newcommand{\ZrOdd}{d~$^3\Phi$}
\newcommand{\ZrOE}{E~$^1\Phi$}
\newcommand{\ZrOee}{e~$^3\Pi$}
\newcommand{\ZrOF}{F~$^1\Delta$}
\newcommand{\ZrOff}{f~$^3\Delta$}
\newcommand{\ZrOsiga}{(1)~$^3\Sigma^+$}

\newcommand{\ZrOX}{X~$^1\Sigma^+$}

\newcommand{\noenergies}{227,118}
\newcommand{\noMaenergies}{13,075}
\newcommand{\noMaenergiesunique}{7,769}
\newcommand{\noEHenergies}{106}
\newcommand{\noCaenergies}{213,937}
\newcommand{\notransitions}{47,662,773}

\defcitealias{Langhoff1990TheoreticalZrO}{L \& B (1990)}
\defcitealias{Huber1979MolecularMolecules}{H \& H (1979)}
\defcitealias{Sorensen2021Near-infraredZrO}{\MOLLIST{}}
\defcitealias{18McKemmishZrO}{\MARVEL{}}
\defcitealias{Hammer1979}{H \& D (1979)}
\defcitealias{Barklem2016}{B \& C (2016)}

\title{Full Spectroscopic Model and Trihybrid Experimental-Perturbative-Variational Line List for ZrO}

\author[A. N. Perri, F. Taher and L. K. McKemmish]
{Armando N. Perri$^{1}$\orcidlink{0009-0005-1737-2569}, Fadia Taher$^{2,3}$\orcidlink{0000-0002-7240-4536}, Laura K. McKemmish$^{1}$\orcidlink{0000-0003-1039-2143}\thanks{E-mail: l.mckemmish@unsw.edu.au}
\vspace*{6pt}\\
$^{1}$School of Chemistry, University of New South Wales, 2052, Sydney, Australia\\
$^{2}$Faculty of Sciences II, Research Platform in Nanosciences and Nanotechnology, Laboratory of Experiments and Computation of \\ Materials and Molecules (EC2M), Lebanese University, Campus Fanar, P.O. Box: 90656, Beirut, Lebanon\\
$^{3}$Faculty of Engineering III, Laboratory of Molecular Quantum Mechanics and Modeling (MQMM), Lebanese University, Hadath \\ Campus, Beirut, Lebanon}

\date{Accepted XXX. Received YYY; in original form ZZZ}

\pubyear{2023} % Enter the current year, for the copyright statements etc.

\begin{document}

\date{\today}

\maketitle

\begin{abstract}
{\noindent 
Zirconium monoxide (ZrO) absorption lines define rare S-type stars and are currently being sought on exoplanets. Successful detection is dependent on an accurate and comprehensive line list, with existing data not ideal for many applications. Specifically, the Plez \etal{} line list is near-complete but has insufficient accuracy for high-resolution cross-correlation, while the Sorensen \& Bernath data  has high accuracy but only considers a small number of spectral bands. 
This article presents a novel spectroscopic model, variational line list and trihybrid line list for the main \ZrO{} isotopologue, as well as isotopologue-extrapolated hybrid line lists for the \isoa{}, \isob{}, \isoc{}, \isod{}~and \isoe{} isotopologues. These were constructed using \DUO{} based on icMRCI-SD/CASSCF~\abinitio{} electronic data calculated using \MOLPRO{}, experimental energies obtained from a previous \Marvel{} data compilation and perturbative energies from Sorensen \& Bernath. The new \ZrO{} \EXOMOL{}-style trihybrid line list, \LLname{}, comprises \noenergies{} energies (\noMaenergies{} experimental) and \notransitions{} transitions up to 30,000~\cm{} (333~nm) between ten low-lying electronic states (\ZrOX{}, \ZrOaa{}, \ZrOA{}, \ZrObb{}, \ZrOB{}, \ZrOC{}, \ZrOdd{}, \ZrOee{}, \ZrOff{} and \ZrOF{}).
The inclusion of experimental energy levels in \LLname{} means ZrO will be much easier to detect using high-resolution ground-based telescopes in the 12,500 -- 17,500~\cm{} (571 -- 800~nm) spectral region. The inclusion of variational energy levels means that the \LLname{} line list has very high completeness and can accurately model molecular absorption cross-sections even at high temperatures. The \LLname{} data will hopefully facilitate the first detection of ZrO in the atmosphere of a hot Jupiter exoplanet, or alternatively more conclusively exclude its presence.}
\end{abstract}


\section{Introduction}

Zirconium monoxide (ZrO) is a second-row transition metal diatomic molecule of astronomical significance. Notably, the strong absorption lines of ZrO are the characteristic spectral signature of rare S-type stars \citep{Keenan1954ClassificationStars}, which~replace the prominent titanium monoxide (TiO) absorption lines of common M-types stars. It is known that S-type~stars have~an atmospheric carbon-to-oxygen ratio near unity and exhibit slow neutron-capture processes (s-processes) that~lead to an overabundance of zirconium \citep{Piccirillo1980,VanEck2017AProperties}. ZrO is also faintly detectable in SC-type stars \citep{Keenan1980SPECTRALSYSTEM}, M-type stars \citep{Bobrovnikoff1934ZrOSTARS} and sunspots \citep{Richardson19311931PASP___43___76R}. There has been some search for ZrO in the atmospheres of hot Jupiter exoplanets without success \citep{Tabernero2021ESPRESSOB,Borsa2021AtmosphericESPRESSO}. 
All current ZrO detections have been made using mid-resolution techniques from either space-based or ground-based telescopes. The former space-based telescopes are particularly useful for the detection of fainter objects, such as hot Jupiter exoplanets, and are promising in the new era of the James Webb Space Telescope (JWST). Despite this, high-resolution cross-correlation (HRCC) techniques using ground-based telescopes have become a critical technique over~the last decade to detect new molecules in exoplanetary atmospheres \citep{Brogi2018RetrievingSpectroscopy,birkby2018exoplanet}, where the Very Large Telescope (VLT) is often employed in combination with ultra-high precision spectrographs such as the most recent %initially with the CRIRES instrument, then HARPS  though more recent studies prefer the 
Echelle SPectrograph for Rocky Exoplanets and Stable Spectroscopic Observations (ESPRESSO) instrument. Cross-correlation is an effective method of signal filtration that facilitates the unambiguous astronomical detection of a molecule. This relies on high-resolution ground-based measurements that are compared to high-resolution model templates simulated from atomic and molecular line lists of sub \cm{} accuracy. High-resolution cross-correlation spectroscopy has been used to successfully identify many diatomic molecules, such as CO \citep{Brogi2018RetrievingSpectroscopy}, OH \citep{Landman2021DetectionWASP-76b} and arguably TiO \citep{Nugroho2017}, in exoplanetary atmospheres. In order to achieve the required accuracy for such detections, it has been shown~that experimentally-derived energies must be explicitly incorporated into a line list \citep{19McKemmishTiO,Syme2021FullCN}. This is especially true for complex transition metal diatomic systems such as ZrO \citep{16TennysonDiatomics}.

The current inability to detect ZrO in an exoplanetary atmospheres using high-resolution cross-correlation is not unexpected given the current state of line list data availability for ZrO. Specifically, for ZrO,~the most complete line list available is from Plez \citep{Plez2003,VanEck2017AProperties}, but this does not contain experimentally-derived energy levels and thus will not have the sub-\cm{} accuracy in the strong lines needed for this technique. The more accurate \Mollist{}~line list \citep{Bernath2020MoLLIST:Spectra,Sorensen2021Near-infraredZrO} only considers the \ZrOB{}~--~\ZrOX{} rovibronic and \ZrOX{} rovibrational transitions. This data  thus misses many strong transitions, significantly increasing the required signal-to-noise for a molecular detection and  greatly restricting the spectral range that can be used. % and is unsuitable for use in all but the most carefully selected spectral regions.  % thus unsuitable for much of the spectral region. in high-resolution cross-correlation spectroscopic studies 

To enable effective use by astronomers with all modern techniques, a line list with both high accuracy for all strong lines positions and high completeness across temperature and spectral ranges is required. For ZrO, this is achieved using the approach outlined by \cite{Syme2021FullCN} for the cyano radical (CN). Specifically, high accuracy is obtained with the high-resolution experimental \Marvel{} energies from \cite{18McKemmishZrO} complemented by a small number of interpolated perturbative (effective Hamiltonian) energies from \cite{Sorensen2021Near-infraredZrO}. High completeness can be created from a new spectroscopic model (fitted to experimental data) constructed herein that yields a variational line list. Finally, the experimental, perturbative and variational energies are combined to generate a novel trihybrid line list for \ZrO{}, named \LLname{}. Various isotopologue energies are predicted from this main line list using standard techniques, where $^{90}$Zr, $^{91}$Zr, $^{92}$Zr, $^{93}$Zr, $^{94}$Zr and $^{96}$Zr have an isotopic abundance of 51.5, 11.2, 17.1, 0.0, 17.4 and 2.8\%, respectively \citep{Nomura1983}.

This article is structured as follows; an overview of the literature experimental, perturbative and \abinitio{} spectroscopic data are presented in Section~\ref{s:abinitio_ZrO}. The \abinitio{}~data~calculated by \cite{Tabet2019TheMonoxide} are also adapted. In Section~\ref{s:ZrOSM}, the spectroscopic model for the main \ZrO{} isotopologue is constructed. In Section~\ref{s:llc}, variational line lists are generated for the \ZrO{}, \isoa{}, \isob{}, \isoc{}, \isod{} and \isoe{} isotopologues, as well as trihybrid and isotopologue-extrapolated hybrid line lists. In Section~\ref{s:lla}, various partition functions, lifetimes and cross sections of ZrO are analysed.

\section[\ZrO{} Spectroscopic Data]{\ZrObold{} Spectroscopic Data}\label{s:abinitio_ZrO}

\begin{figure}
\centering
\includegraphics[width=\linewidth]{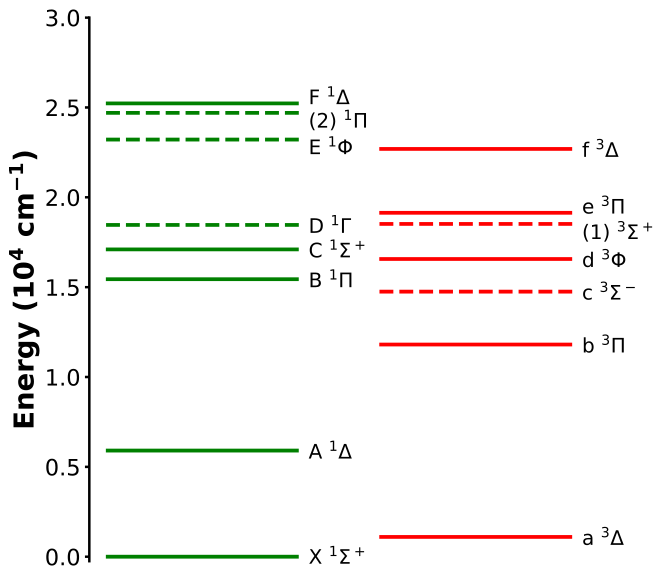}
\caption{The electronic states of ZrO below 25,000~\cm{} with term energies $T_e$ from \protect\cite{18McKemmishZrO} where available and from the MRCI calculations of \protect\cite{Tabet2019TheMonoxide} otherwise. The solid lines indicate electronic states considered in this work.}
\label{f:ZrO_states}
\end{figure}

The rovibronic spectroscopy of ZrO is characterised by many low-lying electronic states as shown in Figure~\ref{f:ZrO_states}. This leads to a complex band structure with intense transitions arising from the singlet and triplet manifolds. From the Boltzmann distribution, the \ZrOX{} and \ZrOaa{} electronic states have significant population such that the \ZrOB{} -- \ZrOX{}, $\alpha$ (\ZrOff{} -- \ZrOaa{}), $\beta$ (\ZrOee{} -- \ZrOaa{}), $\gamma$ (\ZrOdd{} -- \ZrOaa{}) transitions dominate.

\subsection{Experimental Data}
In the literature, the main \ZrO{} isotopologue has been~investigated through many experimental spectroscopic studies. This was recently reviewed and collated in \cite{18McKemmishZrO}, which obtained assigned rovibronic transitions from twelve publications derived from both laboratory measurements and astronomical spectra. Using the \MARVEL{} \citep{07FurtenbacherMARVEL} algorithm, the 22,549 rovibronic transitions validated were inverted to give a set of empirical energies for rovibronic states belonging to the \ZrOX{}, \ZrOA{}, \ZrOB{}, \ZrOC{}, \ZrOF{}, \ZrOaa{}, \ZrObb{}, \ZrOdd{}, \ZrOee{} and \ZrOff{} electronic states. These energies form one self-consistent spectroscopic network containing 8,088 rovibronic states. Apart from unresolved $\Lambda$-doubling in several electronic states, all fine structure is characterised in accordance with Hund's coupling case (a).

\subsection{Perturbative Line Lists}
Beyond the experimental data collated by \cite{18McKemmishZrO}, there exists only perturbative line lists for ZrO~in~the literature. The perturbative (or traditional) methodology relies on an effective Hamiltonian, where experimental data are fit to polynomial expressions through the optimisation of spectroscopic parameters for each electronic state. This approach allows for the prediction of unobserved transitions with \abinitio{} dipole moment curves. This provides a natural extension to experimental data, where the associated line lists interpolate experimental energies to high accuracy. Due to corrections obtained through perturbation theory, however, perturbative line lists typically extrapolate poorly, especially to higher vibrational levels. For this reason, they~are often incomplete as they are limited by experimental data. %Intensities can be readily predicted in the approach based on \abinitio{} dipole moment curves can be used. % as demonstrated by \cite{Sorensen2021Near-infraredZrO} for ZrO. %  In~the literature, the \MOLLIST{} database \citep{Bernath2020MoLLIST:Spectra} is a source of perturbative line lists generated using \PGOPHER{} \citep{Western2017PGOPHER:Spectra}.

The most accurate ZrO perturbative spectroscopic line list in the literature was generated by \cite{Sorensen2021Near-infraredZrO} for the \ZrOB{}~--~\ZrOX{} rovibronic and \ZrOX{} rovibrational transitions. This perturbative analysis relied on a high temperature laboratory emission spectrum in combination with the \abinitio{} \ZrOB{}~--~\ZrOX{} off-diagonal dipole moment curve calculated by \cite{Langhoff1990TheoreticalZrO}.~For reasonable completeness, however, astronomers are currently limited to the perturbative line list constructed by Plez \citep{Plez2003,VanEck2017AProperties}. This is the most comprehensive line list with transition intensities currently available for ZrO (superseded by this work), which was constructed in accordance with the TiO line list methodology of \cite{Plez1998} using spectroscopic parameters from a range of sources. In particular, the Plez line list includes the \ZrOB{}~--~\ZrOX{}, \ZrOC{}~--~\ZrOX{}, \ZrOB{}~--~\ZrOA{}, \ZrOE{}~--~\ZrOA{}, \ZrObb{}~--~\ZrOaa{}, $\gamma$ (\ZrOdd{}~--~\ZrOaa{}), $\beta$ (\ZrOee{}~--~\ZrOaa{}) and $\alpha$ (\ZrOff{}~--~\ZrOaa{}) rovibronic transitions for the \ZrO{}, \isoa{}, \isob{}, \isoc{}, \isod{} and \isoe{} isotopologues.

There is no variational line list currently available for ZrO that has been constructed through numerical solution of the nuclear Schrödinger equation. This prevents reliable extrapolation to unobserved rovibrational states.

\subsection{\AbInitio{} Data}\label{s:ab}
\Abinitio{} spectroscopic data have been calculated for many low-lying electronic states of ZrO. An important historical study was conducted by \cite{Langhoff1990TheoreticalZrO} for twelve electronic states using the MRCI/SA-CASSCF level of theory. Recently, an improved study was performed by \cite{Tabet2019TheMonoxide} for twenty-four electronic states using a larger active space in the CASSCF calculations.

In this work, the \abinitio{} study of \cite{Tabet2019TheMonoxide} was adapted to consider larger bond lengths and explicitly calculate all off-diagonal dipole moment curves. These novel calculations were performed in \Molpro{} 2020.1 \citep{Werner2020ThePackage} using the icMRCI-SD level of theory with reference orbitals obtained from SA-CASSCF calculations. Using the ($a_1$, $b_1$, $b_2$, $a_2$) irreducible representation of the $C_{2v}$ point group, (4, 2, 2, 1) active orbitals from (Zr: $4d$, $5s$, $5p$), (2, 1, 1, 0) active orbitals from (O: $2s$, $2p$) and (3, 1, 1, 0) closed core orbitals from (Zr: $4s$, $4p$; O: $1s$) were selected. These orbitals were optimised for the \ZrOX{}, \ZrOA{}, \ZrOB{}, \ZrOC{}, \ZrOD{}, \ZrOF{}, \ZrOaa{}, \ZrObb{}, \ZrOcc{}, \ZrOdd{}, \ZrOee{}, \ZrOff{} and \ZrOsiga{} electronic states with equal weighting. For the zirconium atom, a contracted version of the cc-pVTZ-PP basis set was employed in combination with the $2h$ function of the aug-cc-pVQZ-PP basis set \citep{Peterson2007Energy-consistentY-Pd}. The associated ECP28MDF effective core potential was used for the twenty-eight frozen electrons in the $1s^2 2s^2 2p^6 3s^2 3p^6 3d^{10}$ orbitals. The remaining twelve electrons in the $4s^2 4p^6 4d^2 5s^2$ orbitals were explicitly taken into the valence space. For the oxygen atom, a contracted version of the aug-cc-pVQZ basis set \citep{Kendall1992ElectronFunctions} was used in an all electron scheme.

\begin{table*}
\centering
\caption{The fitted variational parameters of the potential energy and diagonal coupling curves for the \ZrOX{}, \ZrOA{}, \ZrOB{}, \ZrOC{} and \ZrOF{} singlet electronic states. Apart from the dimensionless parameters $N_L$ and $N_R$, all parameters are rounded to five significant figures in units of Å for $R_e$, Å$^{-i}$ for the $b_i$ vibrational parameters and \cm{} otherwise.}
\label{t:pecZrOsinglet}
\begin{tabular}{cccccc}
\hline
Parameter & \ZrOX{} & \ZrOA{} & \ZrOB{} & \ZrOC{} & \ZrOF{} \\
\hline
$T_e$ & +0.0000E+00 & +5.8854E+03 & +1.5458E+04 & +1.7099E+04 & +2.5156E+04 \\
$R_e$ & +1.7119E+00 & +1.7243E+00 & +1.7561E+00 & +1.7492E+00 & +1.7652E+00 \\
$A_e$ & +6.3637E+04 & +6.3637E+04 & +6.3637E+04 & +6.3637E+04 & +6.3637E+04 \\
$N_L$ & 1 & 1 & 2 & 1 & 2 \\
$N_R$ & 3 & 3 & 3 & 2 & 2 \\
$b_0$ & +1.7365E+00 & +1.7653E+00 & +1.7569E+00 & +1.8300E+00 & +1.9175E+00 \\
$b_1$ & +8.5932E--02 & +1.2397E--01 & +1.7479E--01 & --7.3237E--02 & +2.0378E--01 \\
$b_2$ & +5.5929E--02 & --7.1092E--01 & +1.2322E--01 & +8.8427E--01 & +1.0000E+00 \\
$b_3$ & --8.8825E--02 & +1.0855E+00 & +1.0925E--01 & -- & -- \\
\vspace{-0.5em} \\
$\lambda_\text{q}$ & -- & -- & +9.3589E--05 & -- & -- \\
\hline
\end{tabular}
\end{table*}

\begin{table*}
\centering
\caption{The fitted variational parameters of the potential energy and diagonal coupling curves for the \ZrOaa{}, \ZrObb{}, \ZrOdd{}, \ZrOee{} and \ZrOff{} triplet electronic states. Apart from the dimensionless parameters $N_L$ and $N_R$, all parameters are rounded to five significant figures in units of Å for $R_e$, Å$^{-i}$ for the $b_i$ vibrational parameters and \cm{} otherwise.}
\label{t:pecZrOtriplet}
\begin{tabular}{cccccc}
\hline
Parameter & \ZrOaa{} & \ZrObb{} & \ZrOdd{} & \ZrOee{} & \ZrOff{} \\
\hline
$T_e$ & +1.4179E+03 & +1.2159E+04 & +1.7194E+04 & +1.9146E+04 & +2.3039E+04 \\
$R_e$ & +1.7343E+00 & +1.7414E+00 & +1.7604E+00 & +1.7570E+00 & +1.7869E+00 \\
$A_e$ & +6.3637E+04 & +6.3637E+04 & +6.3637E+04 & +6.3637E+04 & +6.3637E+04 \\
$N_L$ & 1 & 2 & 1 & 1 & 2 \\
$N_R$ & 3 & 2 & 2 & 2 & 2 \\
$b_0$ & +1.6856E+00 & +1.7536E+00 & +1.7844E+00 & +1.7921E+00 & +1.8178E+00 \\
$b_1$ & +6.5169E--02 & +1.0791E--01 & +1.0543E--01 & --1.9849E--01 & +5.5247E--01 \\
$b_2$ & +3.0025E--01 & +2.7406E--01 & +2.5423E--01 & +9.5112E--01 & +1.0000E+00 \\
$b_3$ & --6.6929E--01 & -- & -- & -- & -- \\
\vspace{-0.5em} \\
$A_\text{SO}$ & +3.1204E+02 & --2.9868E+02 & +6.1309E+02 & --1.1935E+02 & +3.6652E+02 \\
$\lambda_\text{SS}$ & --1.3596E+00 & +1.1238E+01 & +4.1128E+00 & --8.5077E+00 & +3.6586E+00 \\
$BO_\text{rot}$ & +6.7111E--03 & +9.9077E--04 & +1.0405E--02 & +8.3545E--03 & +1.1734E--02 \\
$\gamma_\text{SR}$ & --9.7375E--03 & +2.6961E--01 & +1.4856E+00 & +2.7173E--01 & +2.5983E+00 \\
$\lambda_\text{p2q}$ & -- & +1.0989E--01 & -- & -- & -- \\
$\lambda_\text{opq}$ & -- & --1.7547E+01 & -- & -- & -- \\
$\lambda_\text{q}$ & -- & +1.1512E--04 & -- & -- & -- \\
\hline
\end{tabular}
\end{table*}

\begin{figure}
\centering
\includegraphics[width=\linewidth]{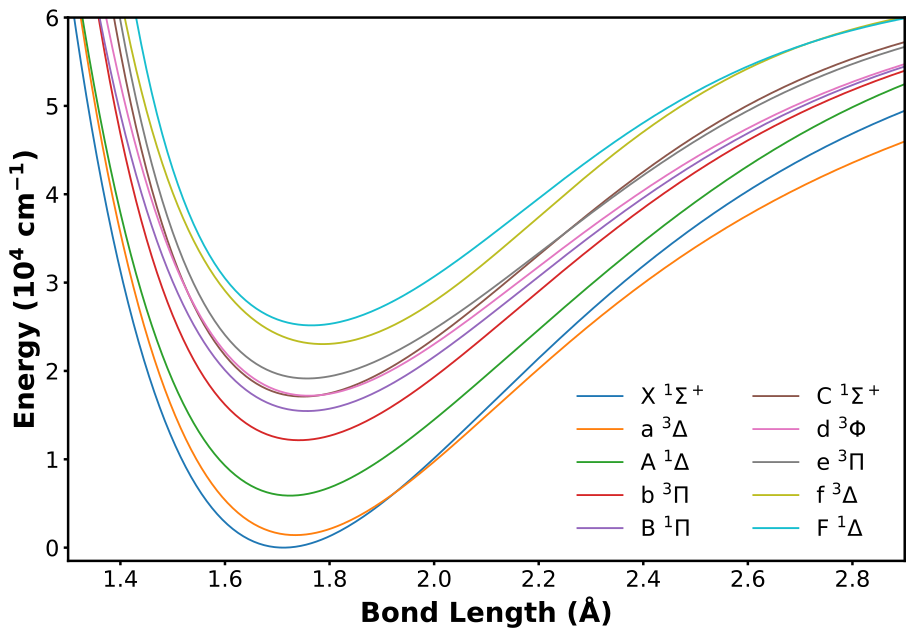}
\caption{The ten ZrO electronic potential energy curves generated in this work. These were modelled using an extended Morse oscillator curve fit to the \MARVEL{} experimental data.}
\label{f:PECZrO}
\end{figure}

\section[\ZrO{} Spectroscopic Model]{\ZrObold{} Spectroscopic Model}\label{s:ZrOSM}
There is currently no available spectroscopic model \citep{21McKemmishDiatomics} (\textit{i.e.} a set of self-consistent potential energy, coupling and dipole moment curves used in the variational model of a molecule) for \ZrO{} in the literature. The absence of this model hinders physically reasonable extrapolation of experimental data to new unobserved transitions. 

Here, the \ZrO{} spectroscopic model is assembled for the ten \ZrOX{}, \ZrOaa{}, \ZrOA{}, \ZrObb{}, \ZrOB{}, \ZrOC{}, \ZrOdd{}, \ZrOee{}, \ZrOff{} and \ZrOF{} electronic states in a \DUO{} \citep{16YurchenkoDuo} \MODEL{} input file. The \ZrOD{}, \ZrOE{} and \ZrOcc{} electronic states were excluded as there is no experimental data verifying their term energies and there are no allowed transitions with the low-lying \ZrOX{} or \ZrOaa{} electronic states. 

\begin{table}
\centering
\caption{A statistical summary of the \ZrO{} absolute energy residuals grouped by electronic state for each vibrational state $v$ in the \MARVEL{} analysis. The root mean square error (RMSE), mean absolute error (MAE) and maximum residual are presented for each vibronic level in units of~\cm{}.}
\label{t:resZrO1}
\begin{tabular}{ccccc}
\hline
State & $v$ & RMSE & MAE & Maximum\\
\hline
\multirow{8}{*}{\ZrOX{}}
& 0 & 1.059E--01 & 9.233E--02 & 2.737E--01 \\
& 1 & 9.139E--02 & 8.528E--02 & 1.762E--01 \\
& 2 & 1.450E--01 & 1.402E--01 & 2.717E--01 \\
& 3 & 1.271E--01 & 1.217E--01 & 2.894E--01 \\
& 4 & 9.591E--02 & 8.155E--02 & 2.546E--01 \\
& 5 & 6.058E--02 & 4.363E--02 & 2.596E--01 \\
& 6 & 1.203E--01 & 1.020E--01 & 2.750E--01 \\
& 7 & 2.229E--01 & 2.131E--01 & 3.361E--01 \\
\vspace{-0.5em} \\
\multirow{6}{*}{\ZrOaa{}}
& 0 & 8.433E--01 & 4.567E--01 & 4.593E+00 \\
& 1 & 8.138E--01 & 6.033E--01 & 4.086E+00 \\
& 2 & 8.516E--01 & 6.148E--01 & 4.115E+00 \\
& 3 & 7.107E--01 & 6.083E--01 & 2.433E+00 \\
& 4 & 9.587E--01 & 8.011E--01 & 1.834E+00 \\
& 5 & 1.395E+00 & 1.105E+00 & 2.423E+00 \\
\vspace{-0.5em} \\
\multirow{2}{*}{\ZrOA{}}
& 0 & 2.576E--01 & 2.115E--01 & 5.842E--01 \\
& 1 & 5.576E--01 & 4.065E--01 & 1.905E+00 \\
\vspace{-0.5em} \\
\multirow{1}{*}{\ZrObb{}}
& 0 & 1.099E--01 & 7.866E--02 & 4.417E--01 \\
\vspace{-0.5em} \\
\multirow{6}{*}{\ZrOB{}}
& 0 & 3.685E--02 & 2.706E--02 & 1.645E--01 \\
& 1 & 3.819E--02 & 2.657E--02 & 1.746E--01 \\
& 2 & 9.147E--02 & 5.152E--02 & 6.875E--01 \\
& 3 & 6.511E--02 & 4.658E--02 & 2.959E--01 \\
& 4 & 5.177E--02 & 3.980E--02 & 1.864E--01 \\
& 5 & 7.217E--02 & 5.803E--02 & 1.988E--01 \\
\vspace{-0.5em} \\
\multirow{1}{*}{\ZrOC{}}
& 0 & 7.253E--02 & 5.394E--02 & 2.160E--01 \\
\vspace{-0.5em} \\
\multirow{5}{*}{\ZrOdd{}}
& 0 & 2.426E+00 & 2.040E+00 & 7.437E+00 \\
& 1 & 1.639E+00 & 1.355E+00 & 5.863E+00 \\
& 2 & 6.627E--01 & 3.900E--01 & 3.734E+00 \\
& 3 & 1.222E+00 & 1.041E+00 & 3.169E+00 \\
& 4 & 3.172E+00 & 2.897E+00 & 5.189E+00 \\
\vspace{-0.5em} \\
\multirow{1}{*}{\ZrOee{}}
& 0 & 8.022E--01 & 5.677E--01 & 3.049E+00 \\
\vspace{-0.5em} \\
\multirow{1}{*}{\ZrOff{}}
& 0 & 1.291E+00 & 9.774E--01 & 3.444E+00 \\
\vspace{-0.5em} \\
\multirow{2}{*}{\ZrOF{}}
& 0 & 3.868E--01 & 3.400E--01 & 8.467E--01 \\
& 1 & 5.116E--01 & 4.338E--01 & 1.001E+00 \\
\hline
\end{tabular}
\end{table}

\subsection{Energy Spectroscopic Model}\label{s:ZrOesm}
The \ZrO{} energy spectroscopic model comprises the potential energy curve of each electronic state, as well as~the coupling and correction curves between these states. Firstly, the potential energy curves for the ten electronic states considered are shown in Figure~\ref{f:PECZrO}, which were modelled using an extended Morse oscillator potential given by
\footnotesize
\begin{equation}\label{e:ZrOemo}
V(R) = T_e + D_e\left[1 - {\rm exp} \left(\sum^{N}_{i=0} b_i \left( \frac{R^4-R^4_e}{R^4+R^4_e} \right)^i \left(R_e-R\right)\right) \right]^2
\end{equation} %\left(A_e - T_e \right)
\normalsize
where $T_e$ is the term energy, $D_e$ is the dissociation energy, $R_e$ is the equilibrium bond length and~$\{b_i$\} are vibrational~fitting parameters. The number of these fitting parameters differs on the left and right of the equilibrium bond length as specified by the parameters $N_L$ and $N_R$ in the \MODEL{} input file, respectively. The dissociation asymptote $A_e=T_e+D_e$ was set to 7.89~eV = 63,637~\cm{} \citep{Murad1975ThermochemicalZrO2} for all ten electronic states.

\begin{table*}
\centering
\caption{The \ZrO{} equilibrium spectroscopic parameters from selected sources reported without error estimates in units of Å for $R_e$ and \cm{} otherwise. The theoretical parameters from \protect\cite{Langhoff1990TheoreticalZrO} are obtained from the MRCI calculations.}
\label{t:newparametersZrO}
\begin{tabular}{cccccccc}
\hline
State & Parameter & \LLname{} & \citetalias{Huber1979MolecularMolecules} & \citetalias{18McKemmishZrO} & \citetalias{Sorensen2021Near-infraredZrO} & \citetalias{Langhoff1990TheoreticalZrO} \\ 
\hline
\multirow{7}{*}{\ZrOX{}} 
& $A_e$ & 63,637 & 63,315 & -- & -- & -- \\
& $R_e$ & 1.7119 & 1.7116 & -- & 1.71199404 & 1.721 \\
& $\omega_e$ & 976.16 & 969.76 & 976.44 & 976.40869 & 979.1 \\
& $\omega_ex_e$ & 3.3758 & 4.90 & 3.45 & 3.44153 & -- \\
& $B_e$ & 0.42361 & 0.42263 & 0.42361 & 0.4235669 & -- \\
& $D_e$ $(10^{-7})$ & 3.19 & 3.19 & 3.19 & -- & -- \\
& $\alpha_e$ $(10^{-3})$ & 1.9572 & 2.3 & 1.97 & 1.9186 & -- \\
\vspace{-0.5em} \\
\multirow{8}{*}{\ZrOB{}} 
& $T_e$ & 15,458 & 15,443 & 15,441.70 & -- & 15,168 \\
& $A_e$ & 63,637 & -- & -- & -- & -- \\
& $R_e$ & 1.7561 & 1.75832 & -- & 1.75631230 & 1.756 \\
& $\omega_e$ & 859.35 & 859 & 859.59 & 859.424252 & 899.9 \\
& $\omega_ex_e$ & 2.9669 & 3 & 2.99 & 2.921697 & -- \\
& $B_e$ & 0.402576 & 0.40154 & 0.40246 & 0.402460322 & -- \\
& $D_e$ $(10^{-7})$ & 3.53 & 3.52 & 3.50 & -- & -- \\
& $\alpha_e$ $(10^{-3})$ & 1.9092 & -- & 1.90 & 1.8770 & -- \\
\hline
\end{tabular}
\end{table*}

The ten potential energy curves were fit to 8,083 unique \MARVEL{} experimental energies \citep{18McKemmishZrO}, where the inverse square of each \MARVEL{} uncertainty was applied as the weighting factor in the fitting procedure. The degenerate \MARVEL{} rovibronic states with unresolved $\Lambda$-doubling were duplicated with opposite parity for completeness. The \abinitio{} potential energy curves from Section~\ref{s:ab} were used to further confine the fit. The fitted parameters for each potential energy curve are provided in Tables~\ref{t:pecZrOsinglet}~and~\ref{t:pecZrOtriplet}.

Additionally, all spin-orbit ($A_\text{SO}$), spin-spin ($\lambda_\text{SS}$), spin-rotation ($\gamma_\text{SR}$), $\Lambda$-doubling ($\lambda_\text{opq}$, $\lambda_\text{p2q}$ and $\lambda_\text{q}$), angular momentum ($L_+$) and Born-Oppenheimer breakdown (BO$_\text{rot}$) interactions were considered in this energy spectroscopic model. A constant value was utilised for these interactions across all bond lengths. For the off-diagonal spin-orbit and angular momenta curves, this constant value was set as its \abinitio{} value at 1.71 Å, which is the equilibrium bond~length of the ground \ZrOX{} electronic state. The remaining values were fit using the \MARVEL{} experimental data. The diagonal coupling parameters are provided in Tables~\ref{t:pecZrOsinglet} and \ref{t:pecZrOtriplet} for all ten electronic states. The off-diagonal coupling parameters can be found in the \DUO{} \MODEL{} file.

Ultimately, the energy spectroscopic model constructed herein can accurately reproduce the \MARVEL{} experimental energies for all ten electronic states. This is shown by the absolute energy residuals summarised statistically in \Cref{t:resZrO1}, which were calculated as the absolute difference between the \MARVEL{} experimental and variational energies.

In \Cref{t:newparametersZrO}, the \ZrOX{} and \ZrOB{} equilibrium spectroscopic parameters from this \ZrO{} model are compared to selected literature studies. There is a general agreement across all sources, with strong agreement to the \MARVEL{} experimental parameters \citep{18McKemmishZrO} as expected from the fitting procedure. The greatest deviations are observed in comparison to the theoretical parameters \citep{Langhoff1990TheoreticalZrO} as expected due to the inaccuracies of quantum chemistry in modelling complex transition metal diatomic molecules \citep{16TennysonDiatomics}.

\subsection{Intensity Spectroscopic Model}\label{s:ZrOism}
\begin{figure}
\centering
\includegraphics[width=\linewidth]{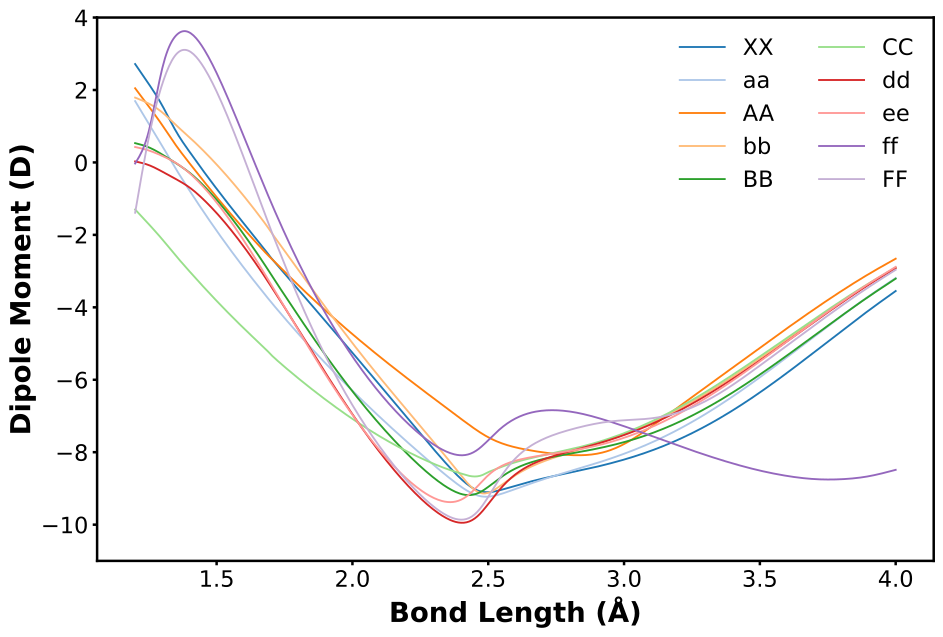}
\caption{The \ZrO{} diagonal (permanent) dipole moment curves as obtained from \abinitio{} calculations.}
\label{f:ZrO_dipole_z}
\end{figure}

The \ZrO{} intensity spectroscopic model contains only \abinitio{} dipole moment curves as is standard practice. These curves were inserted into the \MODEL{} file without fitting to a functional form, where all interpolation was performed by \DUO{} using natural splines. The intensity spectroscopic model includes all diagonal (permanent) and off-diagonal~(transition) dipole moment curves shown in Figures \ref{f:ZrO_dipole_z} and \ref{f:ZrO_dipole_x}, respectively. The latter also includes values calculated by \cite{Langhoff1990TheoreticalZrO} for various off-diagonal curves.

The \abinitio{} dipole moment curves in Figures~\ref{f:ZrO_dipole_z}~and~\ref{f:ZrO_dipole_x} are exceptional for a transition metal diatomic molecule.~Importantly, all curves are smooth and well-defined around the equilibrium bond length of each electronic state. Apart~from the \ZrOff{} electronic state, all diagonal dipole moment curves follow the same pattern as they tend towards zero. The \ZrOff{} diagonal curve likely approaches zero beyond 5~Å. Similarly, the off-diagonal dipole moment~curves generally converge to zero. These off-diagonal curves are consistent with the values calculated by \cite{Langhoff1990TheoreticalZrO} around~the equilibrium bond length. The off-diagonal dipole moment curves associated with the \ZrOff{} electronic state again likely tend towards zero beyond 5~Å. The high \ZrOff{} dipole moment values around 4~Å may be due to a remaining predominant ionic behaviour before reaching the neutral fragment at the dissociation. Nevertheless, it was found that the vibrational basis functions generated by \DUO{} have negligible amplitude beyond 2.5~Å for all electronic states. This eliminates the need for these \abinitio{} curves beyond this bond length.

\begin{table*}
\caption{An extract from the \ZrO{} \STATES{} file. The total file is available at \url{www.exomol.com} and in the supplementary material.}
\label{tab:states}
\resizebox{\textwidth}{!}{
\begin{tabular}{cccccccccccccccccc}
\hline
$n$& $\tilde{E}$ & $g_{\rm tot}$ & $J$ & unc & $\tau$ & $g$ & $p_{+/-}$ & $p_\text{e/f}$ & State & $v$ & $\Lambda$ & $\Sigma$ & $\Omega$ & Source & $\tilde{E}_\text{Ca}$\\
\hline
5633 & 18746.741719 & 13 & 6 & 9.2000E+00 & 2.3475E--03 & 9.4090E--02 & - & f & a3Delta & 20 & -2 & 0 & -2 & Ca & 18746.741719 \\
5634 & 18778.561300 & 13 & 6 & 5.8000E--02 & 6.4538E--08 & 2.3798E--02 & - & f & B1Pi & 4 & -1 & 0 & -1 & EH & 18778.515781 \\
5635 & 18817.703605 & 13 & 6 & 1.2000E--02 & 3.9329E--08 & 2.1485E--01 & - & f & d3Phi & 2 & -3 & 0 & -3 & Ma & 18817.782124 \\
\hline
\end{tabular}
}
\end{table*}

\begin{table}
\centering
\caption{The column descriptors of the ZrO \STATES{} files. Here,~$\hat{\mathbf{L}}$ is the electronic orbital angular momentum, $\hat{\mathbf{S}}$ is the electronic spin angular momentum and $\hat{\mathbf{J}}$ is the total angular momentum~excluding nuclear spin angular momentum \citep{16TennysonDiatomics}.}
\label{tab:statescd}
\begin{tabular}{ccc}
\hline
Column & Symbol & Descriptor \\
\hline
1 & $n$ & State Index \\
2 & $\tilde{E}$ & Energy (in \cm{}) \\
3 & $g_\text{tot}$ & Total Degeneracy \\
4 & $J$ & Total Angular Momentum \\
5 & unc & Uncertainty (in \cm{}) \\
6 & $\tau$ & Lifetime (in s) \\
7 & $g$ & Landé $g$-Factor \\
8 & $p_{+/-}$ & Total Parity \\
9 & $p_\text{e/f}$ & Rotationless Parity \\
10 & State & Electronic State\\
11 & $v$ & Vibrational State \\
12 & {$\Lambda$} & $\hat{\mathbf{L}}$ Projection Along Internuclear Axis \\
13 & {$\Sigma$} & $\hat{\mathbf{S}}$ Projection Along Internuclear Axis \\
14 & {$\Omega$} & $\hat{\mathbf{J}}$ Projection Along Internuclear Axis \\
& & (with $\Omega$ = $\Lambda$ + $\Sigma$) \\
15 & Source & State Source (Ma, EH or Ca) \\
16 & $\tilde{E}_\text{Ca}$ & Variational Energy (in \cm{}) \\
\hline
\end{tabular}
\end{table}

\begin{table}
\caption{An extract from the \ZrO{} \TRANS{} file. The total file is available at \url{www.exomol.com} and in the supplementary material.}
\label{tab:trans}
\centering
\begin{tabular}{ccc}
\hline
$f$ & $i$ & $A_{fi}$ \\
\hline
391 & 1 & 4.7878E--07 \\
392 & 1 & 9.9341E+00 \\
393 & 1 & 2.5024E--02 \\
\hline
\end{tabular}
\end{table}

\begin{table}
\centering
\caption{The column descriptors of the ZrO \TRANS{} files.}
\label{tab:transcd}
\begin{tabular}{ccc}
\hline
Column & Symbol & Descriptor \\
\hline
1 & $f$ & Final State Index \\
2 & $i$ & Initial State Index \\
3 & $A_{fi}$ & Einstein A Coefficient (in s$^{-1}$) \\
\hline
\end{tabular}
\end{table}

\section{\ZrOl{} Line List Construction}\label{s:llc}
\subsection{Methodology Overview}\label{s:mo}
The line list construction approach of \cite{Syme2021FullCN} is followed herein for ZrO, where the \MARVEL{} experimental, \MOLLIST{} perturbative and \DUO{} variational energies are collated into a trihybrid line list. This methodology exploits the advantages of each methodology to produce the most accurate and comprehensive line list for ZrO.

In accordance with \EXOMOL{} line lists conventions, a trihybrid line list is prepared by updating the energies of the variational \STATES{} file. The trihybrid methodology prioritises the experimental energies due to their superior accuracy and reliable uncertainties. These replace the variational energies wherever possible with subsequent interpolation by the perturbative energies. The final trihybrid \STATES{} file includes the methodology and variational energy of each state, where the abbreviations Ma, EH and Ca represent the experimental (\MARVEL{}), perturbative (effective Hamiltonian) and variational (calculated) methodologies, respectively.

\begin{figure}
\centering
\includegraphics[width=\linewidth]{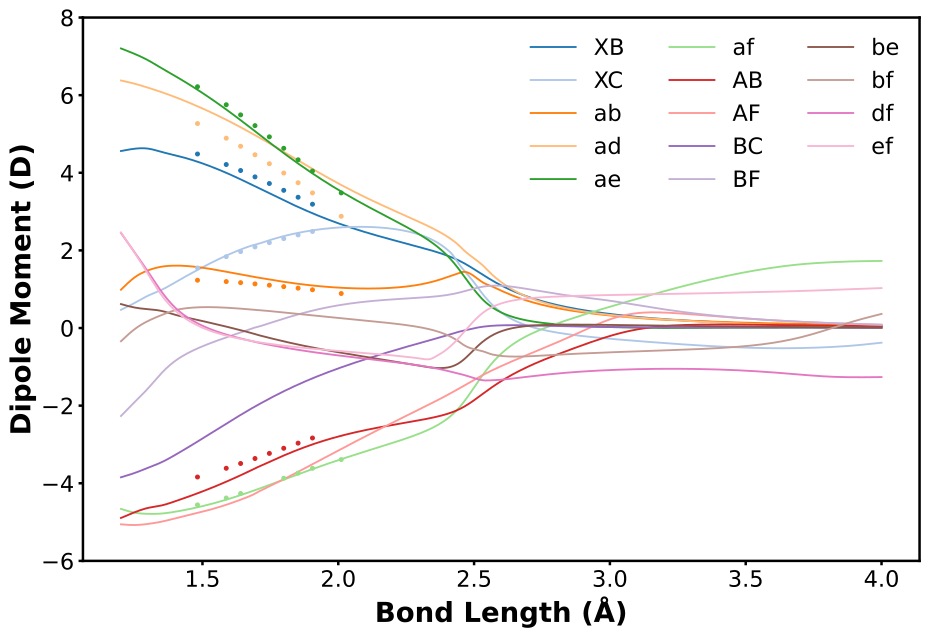}
\caption{The \ZrO{} (transition) off-diagonal dipole moment curves as obtained from \abinitio{} calculations, which were transformed into the Duo $\Lambda$-representation. The coloured dots~indicate values calculated by \protect\cite{Langhoff1990TheoreticalZrO} in the Cartesian representation, where the $\Delta$~--~$\Pi$ and $\Phi$~--~$\Delta$ curves were scaled by $\sqrt{2}$ to convert between real and complex orbitals.}
\label{f:ZrO_dipole_x}
\end{figure}

The \ZrO{} variational line list is output directly from \DUO{} using the energy and intensity spectroscopic model detailed in Section~\ref{s:ZrOSM}. The quantum numbers are given in terms of Hund's coupling case (a). The \ZrO{} trihybrid line list was created by replacing the variational energies with \MARVEL{} experimental energies \citep{18McKemmishZrO} where available, followed by interpolation \MOLLIST{} perturbative energies \citep{Sorensen2021Near-infraredZrO} to cover any missing experimental data. These energy replacements were performed using an external Python script by grouping equivalent rovibronic states from each source by their electronic state, parity, $J$, $v$ and $\Omega$ quantum numbers. % then removing states in accordance with the trihybrid methodology. %This approach enables the calculated variational line list to provide completeness, while the \Marvel{} and perturbative data provide accuracy

For all other isotopologues, a variational line list can~be constructed trivially by changing the nuclear masses in the main spectroscopic model. Following \cite{Polyansky2017ExoMolH217O}, the accuracy of this line list can be improved by applying an isotopologue extrapolation term that shifts each energy based on the difference between the variational and trihybrid line lists of the main \ZrO{} isotopologue as
\begin{equation}
\tilde{E}^\text{iso}_\text{IE} = \tilde{E}^\text{iso}_\text{Ca} + \left(\tilde{E}^\text{main}_\text{tri} - \tilde{E}^\text{main}_\text{Ca}\right)
\end{equation}
where $\tilde{E}^\text{main}_\text{Ca}$ is the calculated variational energy of the main isotopologue, $\tilde{E}^\text{main}_\text{tri}$ is its trihybrid energy, $\tilde{E}^\text{iso}_\text{Ca}$ is the calculated variational energy of a given isotopologue and $\tilde{E}^\text{iso}_\text{IE}$ is its isotopologue-extrapolated energy. In other words, it is assumed that the energy residuals between the hybridised and variational line lists are constant for all isotopologues.

\begin{table*}
\centering
\caption{The partition functions for ZrO as a function of temperature $T$ (in K) from different sources. The left section compares values for the main \ZrO{} isotopologue and the right section gives values for the five other isotopologues considered in this work.}
\label{t:pf}
\resizebox{\textwidth}{!}{
\begin{tabular}{cccc|cccccc}
\hline
$T$ (K) & \ZrO{} & \citetalias{18McKemmishZrO} & \citetalias{Barklem2016} & \isoa{} & \isob{} & \isoc{} & \isod{} & \isoe{} \\					
\hline
0 & 1.00000E+00 & 1.00000E+00 & 1.00000E+00 & 6.00000E+00 & 1.00000E+00 & 6.00000E+00 & 1.00000E+00 & 1.00000E+00 \\
1 & 2.02450E+00 & 2.02446E+00 & 2.02843E+00 & 1.21627E+01 & 2.02970E+00 & 1.21936E+01 & 2.03480E+00 & 2.03970E+00 \\
10 & 1.68071E+01 & 1.68071E+01 & 1.68283E+01 & 1.01007E+02 & 1.68614E+01 & 1.01327E+02 & 1.69137E+01 & 1.69642E+01 \\
100 & 1.64881E+02 & 1.64881E+02 & 1.65280E+02 & 9.90930E+02 & 1.65423E+02 & 9.94124E+02 & 1.65946E+02 & 1.66450E+02 \\
300 & 5.06399E+02 & 5.06325E+02 & 5.07801E+02 & 3.04356E+03 & 5.08103E+02 & 3.05359E+03 & 5.09746E+02 & 5.11330E+02 \\
500 & 1.00696E+03 & 1.00632E+03 & 1.01006E+03 & 6.05271E+03 & 1.01057E+03 & 6.07397E+03 & 1.01405E+03 & 1.01741E+03 \\
800 & 2.51223E+03 & 2.50894E+03 & -- & 1.51030E+04 & 2.52201E+03 & 1.51606E+04 & 2.53144E+03 & 2.54054E+03 \\
1,000 & 4.19424E+03 & 4.18561E+03 & 4.20923E+03 & 2.52168E+04 & 4.21119E+03 & 2.53167E+04 & 4.22754E+03 & 4.24333E+03 \\
1,500 & 1.11864E+04 & 1.10826E+04 & 1.12345E+04 & 6.72634E+04 & 1.12343E+04 & 6.75461E+04 & 1.12806E+04 & 1.13253E+04 \\
2,000 & 2.25679E+04 & 2.18849E+04 & 2.26794E+04 & 1.35711E+05 & 2.26679E+04 & 1.36300E+05 & 2.27645E+04 & 2.28577E+04 \\
3,000 & 6.02447E+04 & 5.32615E+04 & 6.06178E+04 & 3.62304E+05 & 6.05206E+04 & 3.63930E+05 & 6.07869E+04 & 6.10441E+04 \\
5,000 & 2.10516E+05 & 1.36797E+05 & 2.14087E+05 & 1.26600E+06 & 2.11476E+05 & 1.27166E+06 & 2.12401E+05 & 2.13295E+05 \\
\hline
\end{tabular}}
\end{table*}

\subsection{Line List Overview}
The \LLname{} trihybrid \STATES{} file contains \noenergies{} rovibronic states with \noMaenergies{} (\noMaenergiesunique{} unique) experimental states, \noEHenergies{} perturbative states and \noCaenergies{} variational states. These were calculated up to the dissociation limit of 63,637~\cm{}, although the electronic structure of ZrO is poorly understood above 30,000~\cm{} \citep{18McKemmishZrO}. An extract of the final \STATES{} files is shown \Cref{tab:states} with associated column descriptors in \Cref{tab:statescd}.

The \LLname{} trihybrid \TRANS{} file contains \notransitions{} transitions. A maximum transition frequency threshold of 30,000~\cm{} was applied as this range is suitable for most telescopes and ensures that the \ZrOX{} and \ZrOaa{} bands are reliable. These transitions were only calculated for states with lower energy below 20,000~\cm{} given that this contains 99.999\% of the total population at 2,000~K according to the Boltzmann distribution. A vibrational basis set up to $v$~=~30 was used for each electronic state with rotational states calculated to $J$~=~200. An extract of the final \TRANS{} files is shown \Cref{tab:trans} with associated column descriptors in \Cref{tab:transcd}. Additionally, isotopologue-extrapolated hybrid line lists were prepared for \isoa{}, \isob{}, \isoc{}, \isod{} and \isoe{} using the technique in Section~\ref{s:mo}.

\vspace{-6pt}

\subsection{Uncertainties}

The uncertainties assigned to the experimental, perturbative and variational energies were all evaluated differently. 

For the main \ZrO{} isotopologue, the experimental uncertainties were obtained directly from the \MARVEL{} analysis. The perturbative uncertainties were averaged from the experimental uncertainties with the same vibrational and rotational quantum numbers, where the experimental and perturbative energies are known to have similar accuracy when interpolating. The variational uncertainties were approximated using the energy residuals in \Cref{t:resZrO1}. For energies within a vibrational state in the \MARVEL{} analysis, the uncertainty is given as the largest energy residual with the same vibrational quantum number. For energies within a vibrational state unknown by the \MARVEL{} analysis, the uncertainty is given as twice the largest energy residual in the same electronic state. All uncertainties were rounded to two significant figures.

For the five other isotopologues, all uncertainties were doubled to account for errors arising from the isotopologue-extrapolation technique.

\vspace{-6pt}

\section{\ZrOl{} Line List Analysis}\label{s:lla}

\EXOCROSS{} \citep{Yurchenko2018EXOCROSS:Lists} was used to generate partition functions, lifetimes and cross sections for all six ZrO isotopologues studied here.

\subsection{Partition Functions}

The partition functions $Q(T)$ were calculated using
\begin{equation}
Q(T) = \sum_n g_n^\text{ns}(2J_n+1)e^{-c_2\tilde{E}_n/T}
\end{equation}
where $g^\text{ns}$ is the nuclear spin statistical weight factor, $J$ is the total angular momentum excluding nuclear spin, $c_2 = hc/k_B$ (in cm K), $\tilde{E}$ is the energy term value (in cm$^{-1}$), and $T$ is temperature (in K). 

In \Cref{t:pf}, the six partition functions are evaluated at selected temperatures. For \ZrO{}, the partition function is compared to values calculated by \cite{18McKemmishZrO} and \cite{Barklem2016}. There is strong agreement between the \LLname{} data and \cite{Barklem2016}, even at 5,000~K, thus strongly corroborating both data sets. The \Marvel{}-only data are surprisingly complete to 2,000~K, although the lack of vibrational states results in an underestimation of the partition function at higher temperatures. 

\subsection{Lifetimes}
The lifetime $\tau_i$ of each rovibronic state is calculated as 
\begin{equation}
\tau_i = \frac{1}{\sum_f A_{fi}}
\end{equation}
where $A_{fi}$ are Einstein A coefficients. 

In \Cref{t:lifetimes}, selected~excited vibronic lifetime of \ZrO{} are presented as averaged over all rotational states with $J\le10$ with $\abs{\Omega}$-resolution. For three electronic states, the vibronic lifetimes of \LLname{} can be compared to experimental values from \cite{Hammer1979} and \cite{Simard1988}; this comparison provides the best experimental validation of the computed off-diagonal (transition) dipole moments. The \ZrOC{} state agreement is exceptional. The \ZrOB{} and \ZrOee{} state theory-experiment agreement is reasonable, with our new theoretical predictions slightly closer to experimental data than the earlier theoretical calculations.

These vibronic lifetimes are also compared to theoretical values from \cite{Langhoff1990TheoreticalZrO}, with typically strong agreement as expected given the similarity in the dipole moment curves. The largest error is for the \ZrObb{} lifetime, which is unexpected since the dominant contribution, the \ZrObb{}~--~\ZrOaa{} dipole moment curve, is almost identical. This was investigated to ensure the correct conversion from \Molpro{} output to \Duo{} input was performed, but the cause of this discrepancy could not be identified. It should be noted that the \ZrOee{}~--~\ZrOaa{} transition was treated in the same way, resulting in \ZrOee{} lifetimes very similar to \cite{Langhoff1990TheoreticalZrO} and close to experimental data. Any \ZrObb{} experimental lifetimes will thus be helpful here. % to resolve this discrepenacy.

\begin{table}
\centering
\caption{The excited vibronic lifetimes $\tau$ (in ns) of \ZrO{} with $\abs{\Omega}$-resolution averaged over all rotational states with $J\le10$.}
\label{t:lifetimes}
\begin{tabular}{cccccc}
\hline
State & $v$ & $\abs{\Omega}$ & \LLname{} & \citetalias{Langhoff1990TheoreticalZrO} & Experiment \\
\hline
\multirow{3}{*}{\ZrObb{}}
& 0 & 0 & 1580 & 1070$^\text{a}$ & -- \\
& 0 & 1 & 1630 & 1070$^\text{a}$ & -- \\
& 0 & 2 & 1720 & 1070$^\text{a}$ & -- \\
\vspace{-0.5em} \\
\multirow{1}{*}{\ZrOB{}}
& 0 & 1 & 62.9 & 56.4 & 83$^\text{ac}$ \\
\vspace{-0.5em} \\
\multirow{1}{*}{\ZrOC{}}
& 0 & 0 & 130 & 136 & 126(9)$^\text{ac}$ \\
\vspace{-0.5em} \\
\multirow{3}{*}{\ZrOdd{}}
& 0 & 2 & 40.3 & 45.9$^\text{a}$ & -- \\
& 0 & 3 & 38.1 & 45.9$^\text{a}$ & -- \\
& 0 & 4 & 35.8 & 45.9$^\text{a}$ & -- \\
\vspace{-0.5em} \\
\multirow{3}{*}{\ZrOee{}}
& 0 & 0 & 25.5 & 24.4$^\text{a}$ & 32.5(2)$^\text{ab}$\\
& 0 & 1 & 26.4 & 24.4$^\text{a}$ & 32.5(2)$^\text{ab}$\\
& 0 & 2 & 27.3 & 24.4$^\text{a}$ & 32.5(2)$^\text{ab}$\\
\vspace{-0.5em} \\
\multirow{3}{*}{\ZrOff{}}
& 0 & 1 & 21.7 & 28.3$^\text{a}$ & -- \\
& 0 & 2 & 21.8 & 28.3$^\text{a}$ & -- \\
& 0 & 3 & 21.5 & 28.3$^\text{a}$ & -- \\
\hline
\end{tabular}
\centering
\begin{tabular}{l}
\begin{minipage}{\columnwidth}
a No $\abs{\Omega}$-resolution provided. \\
b The experimental lifetime for $v'=0$, $J'=77$ from \cite{Hammer1979}. The equivalent \LLname{} lifetime is $\tau= 27.4$ ns when averaged over all fine structure.\\
c The experimental lifetimes for $v'=0$ from \cite{Simard1988}. \\
\end{minipage}
\end{tabular}
\end{table}

\subsection{Cross Sections}\label{s:ZrOacs}

\begin{figure*}
\centering
\includegraphics[width=\linewidth]{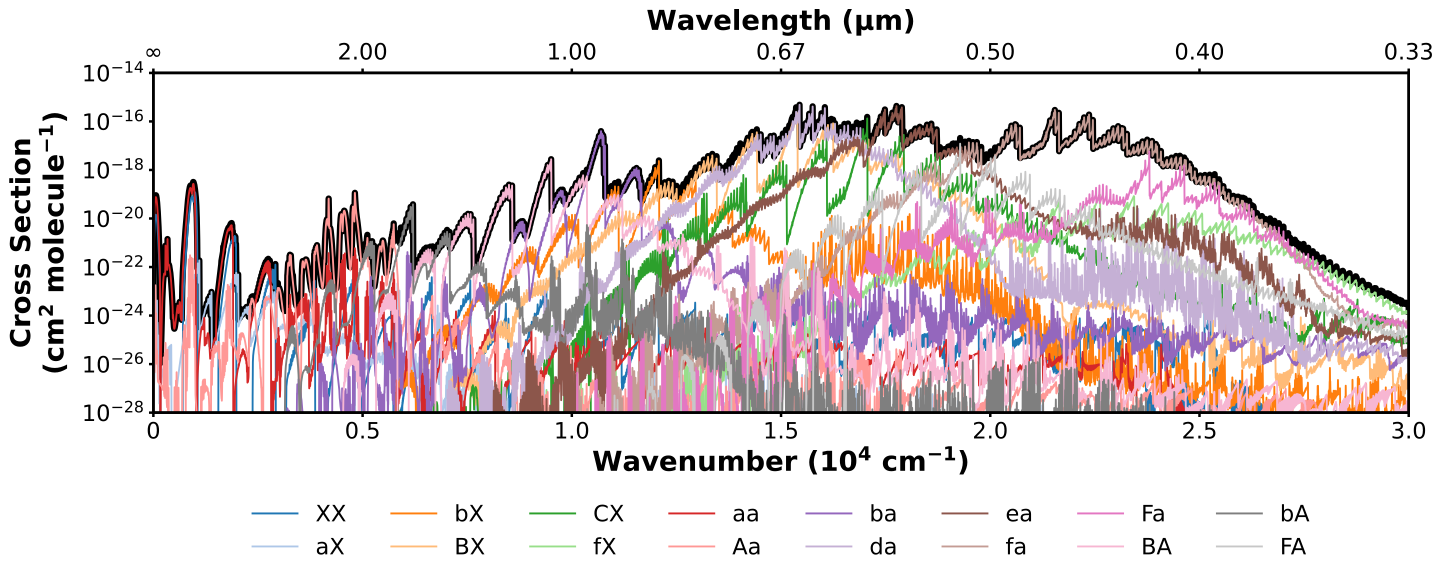}
\caption{The total \ZrO{} absorption cross section decomposed at 2,000~K for 0 -- 30,000~\cm{}.}
\label{f:xsec_all_1_ZrO}
\vspace{-9pt}
\end{figure*}

\begin{figure*}
\centering
\includegraphics[width=\linewidth]{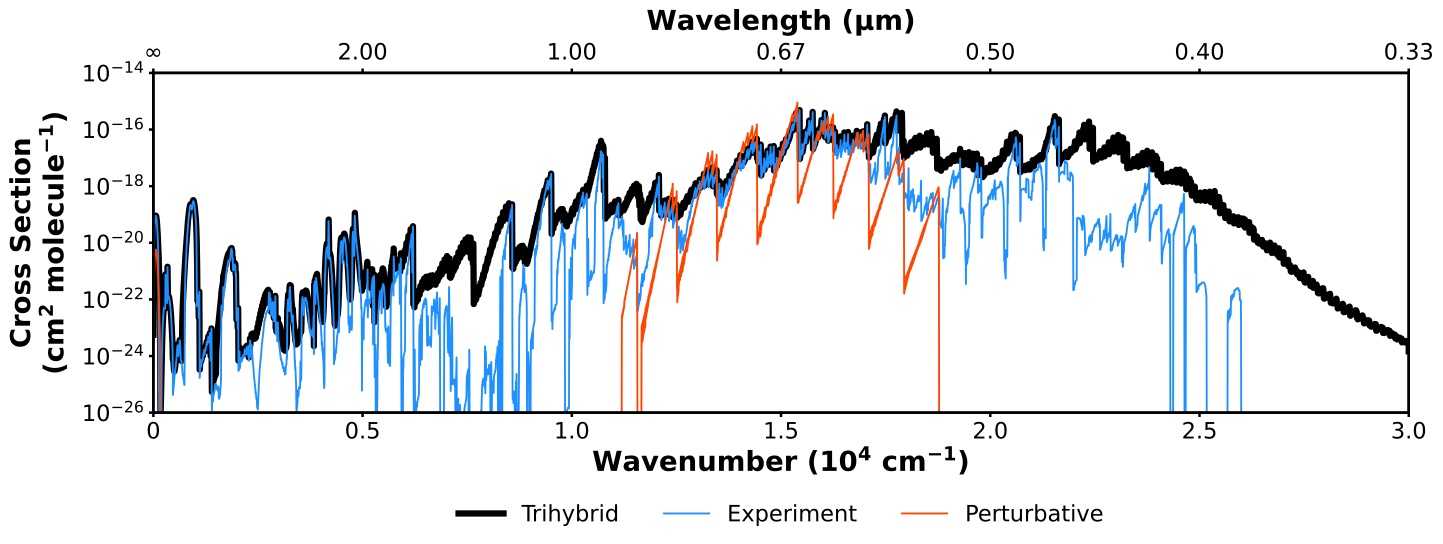}
\caption{The total \ZrO{} absorption cross section simulated at 2,000 K from different sources. The black, blue and orange cross~sections show transitions accounted for by the \LLname{} trihybrid line list, \MARVEL{} experimental line list by \protect\cite{18McKemmishZrO} (with variational intensities) and \MOLLIST{} perturbative line list by \protect\cite{Sorensen2021Near-infraredZrO}, respectively.}
\label{f:xsec_all_3_ZrO}
\vspace{-9pt}
\end{figure*}

\begin{figure*}
\centering
\includegraphics[width=\linewidth]{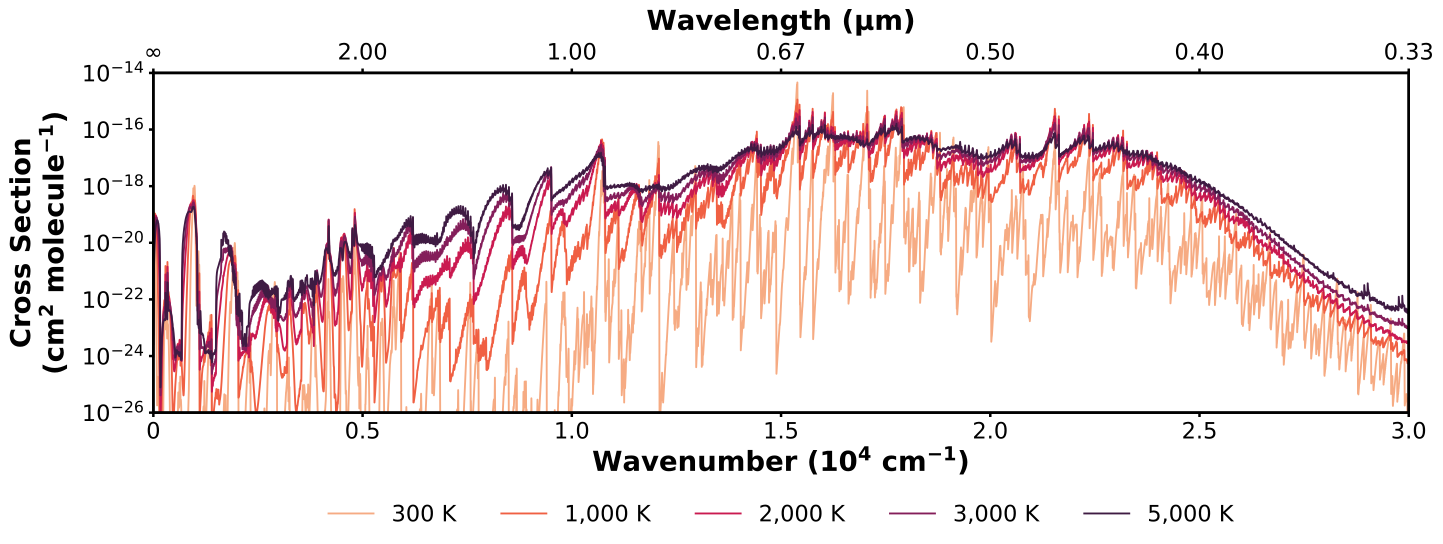}

\caption{The total \ZrO{} absorption cross section simulated at several temperatures.}
\label{f:xsec_all_4_ZrO}
\end{figure*}

\begin{figure*}
\centering
\includegraphics[width=\linewidth]{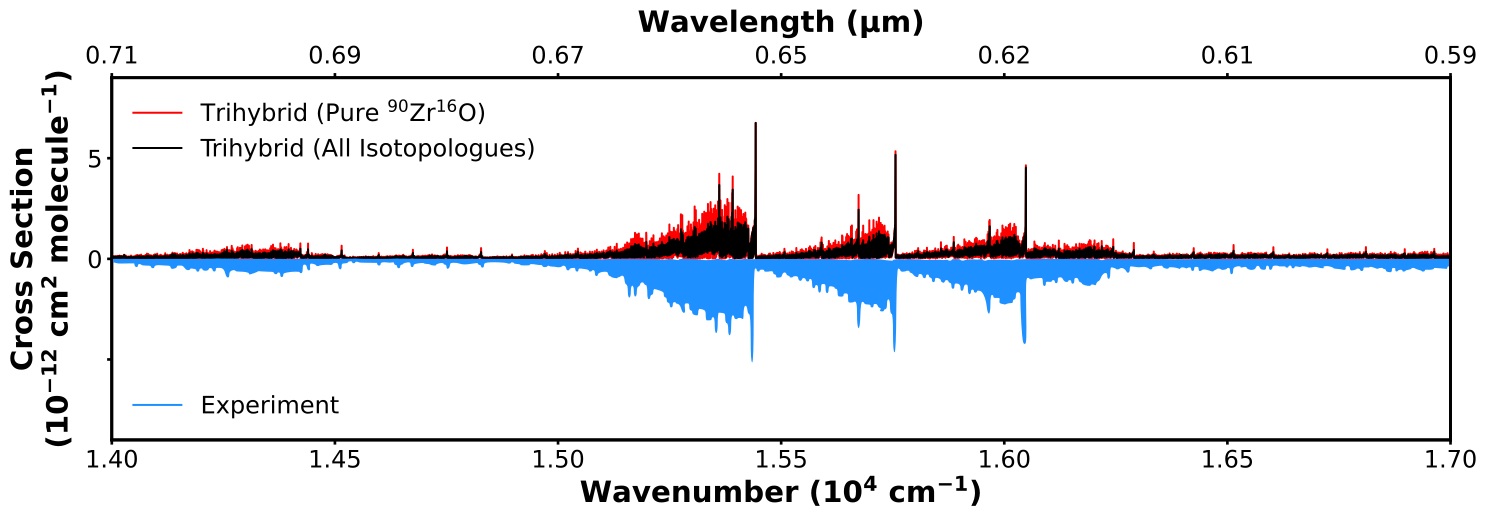}
\caption{A comparison of the ZrO emission cross section simulated from the trihybrid line list with an experimental spectrum at 2,390 K for 14,000 -- 17,000~\cm{}. The red trihybrid cross section was generated using only the main \ZrO{} isotopologue, whereas the black trihybrid cross section was generated using all six isotopologues scaled in accordance with their zirconium isotopic abundance. The experimental cross section was obtained from \protect\cite{Sorensen2021Near-infraredZrO} and copied over~the~trihybrid cross sections.}
\label{f:xsec_all_5_ZrO}
\end{figure*}

Here, cross sections are presented for the main \ZrO{} isotopologue only. The absorption cross sections in Figures \ref{f:xsec_all_1_ZrO}, \ref{f:xsec_all_3_ZrO} and \ref{f:xsec_all_4_ZrO} were simulated using a Gaussian lineshape with a half-width at half-maximum of 2~\cm{}. The emission cross section in Figure~\ref{f:xsec_all_5_ZrO} was simulated using a half-width at half-maximum of 0.1~\cm{} to replicate the laboratory spectrum.

In Figure~\ref{f:xsec_all_1_ZrO}, the total \ZrO{} absorption cross section at 2,000 K is decomposed into its primary transition bands for 0 -- 30,000~\cm{}. The spectrum possesses many intense transitions in all spectral regions. The visible region is the most prominent, where the \ZrOB{} -- \ZrOX{}, $\alpha$ (\ZrOff{} -- \ZrOaa{}), $\beta$ (\ZrOee{} -- \ZrOaa{}), $\gamma$ (\ZrOdd{} -- \ZrOaa{}) transitions dominate.

In Figure~\ref{f:xsec_all_3_ZrO}, the total \ZrO{} absorption cross section at 2,000~K is generated using three line lists. The black, blue and orange cross sections indicate transitions provided by the trihybrid, \MARVEL{} experimental (with variational intensities) and \MOLLIST{} perturbative line lists, respectively. As expected, the trihybrid line list produces the most comprehensive cross section due to its inclusion of the variational transition intensities calculated in this work. The \MARVEL{} analysis offers good coverage in the visible and near-infrared spectral regions corresponding to the major \ZrOB{} -- \ZrOX{}, $\alpha$, $\beta$ and $\gamma$ rovibronic transitions. The \MOLLIST{} cross section is only present for the \ZrOB{} -- \ZrOX{} transition. Evidently, the \MOLLIST{} line list can not sufficiently model the total \ZrO{} absorption cross section given that it does not account for many other overlapping transitions.

In Figure~\ref{f:xsec_all_4_ZrO}, the total \ZrO{} absorption cross section is simulated at several temperatures. The weaker bands in the cross section increase in intensity and become less defined with increasing temperature.

In Figure~\ref{f:xsec_all_5_ZrO}, the \ZrO{} emission cross section simulated from the trihybrid line list is compared with an experimental emission spectrum for 14,000 -- 17,000~\cm{}, which was recorded at 2,390~K with a resolution of 0.04~\cm{} and was used in the perturbative analysis of \cite{Sorensen2021Near-infraredZrO}. As expected, the transition frequencies are in excellent agreement between both spectra, including for many weaker bands. The transition intensities from the trihybrid line list provide an accurate description of the experimental spectrum, which supports the off-diagonal dipole moment curves in Figure~\ref{f:ZrO_dipole_x}. Overall, this comparison is a good improvement over an analogous figure provided by \cite{Sorensen2021Near-infraredZrO} using their \ZrOB{} -- \ZrOX{} line list.

\subsection{High-Resolution Considerations}
In order to assess the suitability of a line list for use in the high-resolution cross-correlation (abbreviated as HRCC) detection of a molecule, it is critical to know the accuracy with which line positions are known. The accuracy of a transition frequency depends on the source of the energies of its upper and lower state. If both the upper and lower state energies are known from \Marvel{} data (with typically low uncertainties), then its frequency can be assumed to be known to high accuracy. Conversely, if one or both of the energy levels are calculated from the Duo variational spectroscopic model, then the frequency will have much higher uncertainty; it is assumed here that it is insufficiently accurate for HRCC studies. The effective Hamiltonian perturbative data can be neglected from this discussion due to the low number of energies in the \LLname{} line list.

Again, in Figure~\ref{f:xsec_all_3_ZrO}, the cross section of the full \LLname{} trihybrid line list (assumed to be near complete based on partition function considerations and the inclusion of all predicted low-lying electronic states) is compared with the cross section predicted solely by \Marvel{} data (assumed to be highly accurate). In spectral regions where the two curves overlap, it should be understood that the transition frequency of the strong line is predicted to experimental (\Marvel{}) accuracy. In this case, Figure~\ref{f:xsec_all_3_ZrO} shows that the \LLname{} trihybrid line list is suitable for HRCC primarily between 12,500 -- 17,500~\cm{} (571 -- 800~nm) at 2,000~K, but not in most other spectral regions. Fortunately, this region is typically quite experimentally accessible. It should~be noted that this figure can be readily generated at different temperatures using ExoCross and examined at higher resolutions to ensure suitability for a desired application. 

Additionally, Figure~\ref{f:xsec_all_3_ZrO} clearly demonstrates the challenges one might face using the previous \cite{Sorensen2021Near-infraredZrO} data alone for HRCC. Narrow spectral windows would be needed, often  undesirable as they reduce signal strength. Further, the  \ZrOdd{}~--~\ZrOaa{} band (not present in this earlier data) overlaps almost entirely with the well-predicted \ZrOB{}~--~\ZrOX{} band;  cross-correlation using the \cite{Sorensen2021Near-infraredZrO} as template data would therefore utilise fewer lines and thus require a much higher signal-to-noise observational data for detection than cross-correlation using our new \LLname{} line list data. %this means fewer line Notably, the well-predicted \ZrOB{}~--~\ZrOX{} bands are too close to other spectral bands unless a very narrow spectral window is selected. Narrow spectral windows are undesirable for HRCC as they reduce signal strength. 

\begin{figure}
\centering
\includegraphics[width=\linewidth]{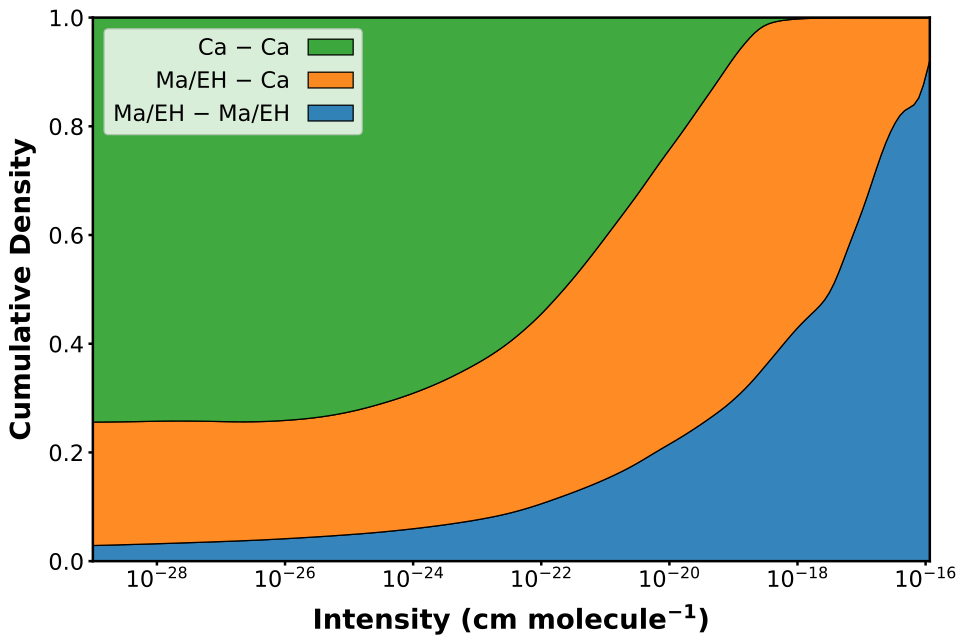}
\caption{The cumulative density of energy sources as a function of transition intensity. The abbreviations Ma, EH and Ca represent the experimental (MARVEL), perturbative (effective Hamiltonian) and variational (calculated) methodologies, respectively.}
\label{f:intdis}
\end{figure}

A useful complementary perspective on high-resolution completeness is obtained from Figure~\ref{f:intdis}. This plot is read~by looking at vertical slices to visualise the source of upper~and lower state energies for transitions above that intensity. For example, the vertical slice at $10^{-18}$ cm molecule$^{-1}$ shows that about half of all transitions with intensity greater than $10^{-18}$ cm molecule$^{-1}$ have both upper and lower energies from the high-resolution \Marvel{}/\MOLLIST{} data; this means that the \LLname{} line list cannot be trusted to accurately predict about half of all transitions with strength above $10^{-18}$ cm molecule$^{-1}$ at 2,000~K. This statistic emphasises the importance of carefully selecting the spectral range for HRCC in ZrO. For molecules with sufficient \Marvel{} data, there is thus expected to be fewer restrictions on the spectral regions suitable for HRCC detections.

In order to improve the high-resolution completeness of the \LLname{} trihybrid line list for HRCC studies, the \ZrO{} \MARVEL{} analysis would need to be expanded with new experimental data. The most important gaps are the \ZrOee{}~and \ZrOff{} electronic states, which will enable accuracy at~shorter wavelengths than currently possible, while the \ZrObb{} electronic state would be useful for longer wavelength studies. %Currently, only the \ZrOX{}, \ZrOB{}, \ZrOaa{} and \ZrOdd{} electronic states possess data for more than two vibrational levels and many spectral bands of moderate intensity have no data. The lack of experimental data for other electronic states significantly limits the fitting procedure, which lowers the accuracy of variational extrapolation.

\section{Conclusions and Future Directions}
Modern searches for molecules like ZrO rely on both traditional mid-resolution space-based observations (such as~from JWST) and very high-resolution ground-based observations (such as from ESPRESSO/VLT). In order to enable the~former and thus successfully model complex physical and chemical processes in astronomical bodies, experience has shown that a high coverage of absorption lines across a wide temperature and spectral range is important \citep{Fortenberry2017QuantumSpectroscopy, Tennyson2012ExoMol:Atmospheres}. In contrast, to enable molecule detection using the modern high-resolution cross-correlation spectroscopic technique, line positions for strong lines must be accurate to about 0.1 \cm{} over the desired spectral region; astronomers should select spectral regions based on the accuracy of the best available line lists.

Therefore, the joint goals of accuracy and completeness are achieved by carefully combining experimental data (to provide high accuracy) with computational models (to provide high completeness). In the construction of the \LLname{} trihybrid line list for ZrO, experimental data are combined with data generated from perturbative (effective Hamiltonians) and variational models. The novel spectroscopic model presented herein considered the ten low-lying \ZrOX{}, \ZrOaa{}, \ZrOA{}, \ZrObb{}, \ZrOB{}, \ZrOC{}, \ZrOdd{}, \ZrOee{}, \ZrOff{} and \ZrOF{} electronic states of ZrO. % A novel spectroscopic model, variational line list and trihybrid line list have been developed for the main \ZrO{} isotopologue, in addition to novel pseudo-hybrid line lists for the 

Through the trihybrid methodology, the \LLname{} line list is now the most accurate and comprehensive~line list available for ZrO. Line lists are available for the main \ZrO{} isotopologue, as well as the \isoa{}, \isob{}, \isoc{}, \isod{} and \isoe{} isotopologues. This new data can be reliably used for molecular detections using high-resolution cross-correlation techniques in the 12,500 -- 17,500~\cm{} (571 -- 800~nm) spectral region, but should not be used at other wavelengths due to the insufficiently complete experimental data. Studies of other vibronic bands of the \ZrOB{}~--~\ZrOX{} and \ZrObb{}~--~\ZrOaa{} transitions would enable use of high-resolution techniques in a wider spectral region.

\section*{Acknowledgements}
This research was undertaken with the assistance of resources from the National Computational Infrastructure (NCI Australia), an NCRIS enabled capability supported~by the Australian Government.

The authors declare no conflicts of interest. 

\section*{Data Availability Statement}
All line list data for this article are available on the ExoMol website, specifically: 
\begin{itemize}[leftmargin=*]
\item The main \LLname{} \MODEL{} file (90Zr16O\_ZorrO.model).
\item The main \LLname{} partition function from 0 -- 10,000 K (90Zr16O\_ZorrO.pf).
\item The main \LLname{} \STATES{} file (90Zr16O\_ZorrO.states).
\item The main \LLname{} \TRANS{} file (90Zr16O\_ZorrO.trans).
\item All other isotopologue \MODEL{}, .pf, \STATES{} and \TRANS{} files (in their respective folders).
\end{itemize}

The \Duo{} spectroscopic model, partition function and \STATES{} files for the main \ZrO{} isotopologue are included as supporting information for this paper. The \MOLPRO{} quantum chemistry input files are also included for the MRCI calculations.

\balance

%\bibliography{ZrO_refs}
%\bibliographystyle{mnras}

\end{document}